# Raman excitation spectroscopy of carbon nanotubes: effects of pressure medium and pressure

A.J. Ghandour,[a] A. Sapelkin,[a] I. Hernandez,[a] D.J. Dunstan,[a]* I.F. Crowe[b] and M.P. Halsall[b]

[a]*School of Physics and Astronomy, Queen Mary University of London, London, United Kingdom*;
[b]*Photon Science Institute and School of Electrical and Electronic Engineering, University of Manchester, Manchester, United Kingdom*

Raman excitation and emission spectra for the radial breathing mode (RBM) are reported, together with a preliminary analysis. From the position of the peaks on the two-dimensional plot of excitation resonance energy against Raman shift, the chiral indices (*m, n*) for each peak are identified.  Peaks shift from their positions in air when different pressure media are added – water, hexane, sulphuric acid – and when the nanotubes are unbundled in water with surfactant and sonication. The shift is about 2 – 3 cm$^{-1}$ in RBM frequency, but unexpectedly large in resonance energy, being spread over up to 100meV for a given peak. This contrasts with the effect of pressure.  The shift of the peaks of semiconducting nanotubes in water under pressure is orthogonal to the shift from air to water.  This permits the separation of the effects of the pressure medium and the pressure, and will enable the true pressure coefficients of the RBM and the other Raman peaks for each (*m*, *n*) to be established unambiguously.

Carbon nanotubes: pressure medium and pressure

---

*Corresponding author. Email: d.dunstan@qmul.ac.uk.

## Introduction

Although the pressure dependence of the Raman spectra of carbon nanotubes has been studied for over fifteen years, there is little agreement between the results of the different laboratories nor is there consistency with simple models [1]. Most studies have concentrated on the G-band, which derives from the graphite $E_{2G}(2)$ band. The graphite band shifts at 4.7cm$^{-1}$GPa$^{-1}$ [2], and while the pressure coefficient of the corresponding band in graphene is consistent with this [3], the reported pressure coefficients of the nanotube band are not. There is no consistent variation according to the nanotube diameter, whether the nanotubes are open (filled) or closed (empty), or whether they are circular or collapsed. Moreover, the pressure coefficients vary unexpectedly with pressure medium and excitation energy [1, 4]. Some clarity is perhaps emerging on the ovalisation and collapse of closed tubes under pressure [5, 6]. Here we focus on the low-frequency radial breathing mode (RBM) at ambient and low pressure (pressures below any ovalisation or collapse).

Since Kataura *et al.* published the eponymous plot, in which each nanotube specified by the chiral numbers (*m*, *n*) corresponds to a point on a plot of excitation resonance energy against diameter [7], a lot of work has been put into theoretical and experimental refinement of the Kataura plot. Maultzsch *et al.* published a version in which the diameter axis is replaced by the inverse of the RBM frequency, and the (*m*, *n*) identifications come from a careful comparison of theory and experiment [8]. Telg *et al.* showed that the identification could be carried over reliably to the anti-Stokes RBM peaks, which have a different resonance energy [9]. Most of this work has been carried out on nanotubes unbundled and dispersed in water by surfactants and sonication; different surfactants cause different shifts in the excitation resonance energy which can be of opposite sign for semiconducting and metallic tubes [8]. Cambré *et al.* [10] showed that the presence of water inside the nanotubes could be detected by the shift in position on the Kataura plot.

We report experiments designed to measure systematically the effects of pressure medium and of pressure on the RBM of tubes identified in this way by their (*m*, *n*) chiral numbers. The objective is to establish whether such experiments can give unambiguous pressure data for different tubes, decoupled from the effects of the pressure medium.

## Experimental Results

Raman spectroscopy was carried out using Ti-sapphire lasers. The excitation energy was varied in small steps from 1.48eV to 1.78eV, and the RBM spectra were recorded from 210cm$^{-1}$ to 305cm$^{-1}$ for Hipco tubes, and from 140cm$^{-1}$ to 185cm$^{-1}$ for Carbolex tubes, for each excitation energy. These RBM frequency ranges were chosen for the diameters of these tubes (Hipco, 0.7 – 1.4nm; Carbolex, 1.2 – 1.6nm), to select a region of the Kataura plot accessible by our lasers and with a moderate density of nanotubes. Preliminary results are shown here; a fuller analysis will be presented elsewhere.

The nanotube samples were measured first without unbundling. RBM peaks and their excitation spectra were well-resolved, as expected from, e.g., the results for bundled tubes presented by Fantini *et al.* [11]. Data were recorded in air with a laser spot about 15μm diameter. The power was increased until at about 55mW effects attributed to a rise in temperature were observed (a redshift of the RBM by about 1cm$^{-1}$, corresponding to a temperature rise of perhaps 70° [12]). The power was then reduced to 20mW and this power was used for all subsequent measurements. For each pressure medium (water, hexane and sulphuric acid), the nanotubes were simply wetted with a few drops of the liquid. For comparison, we also unbundled the nanotubes as a dispersion in water using (following O'Connell *et al.* [13] and many subsequent authors) sodium dodecyl sulphate as surfactant, and sonication for one hour. Unbundling was confirmed by the absence of the Breit-Wigner-Fano peak in the G-band for metallic tubes [14].

Typical spectra are shown in Fig.1. Lorentzian peaks were fitted to the spectra, by an iterative process in which a set of peaks were sought which varied only in intensity between successive spectra and which corresponded in a consistent manner to the RBM frequencies in the Kataura plots of Araujo *et al*. [15, 16] taking into account also the identifications made in experimental work by other authors [e.g. 17]. Uncertainties in the RBM peak frequencies found in this way are about $1 cm^{-1}$. Then the excitation energy for which each peak was most intense was identified (to an accuracy of about 10meV), yielding the co-ordinates for the data points plotted in Fig.2.

Both when bundled and when unbundled, the spectra for the Hipco tubes could be identified as belonging to seven different tubes from three semiconducting families, $2m + n = 25$, 22 and 19 (Figure 2). The solvents all blue-shifted the RBM frequency by $2 – 3 cm^{-1}$ relative to air. The dramatic effect, however, is on the excitation resonance energy, which was up to 100meV different for bundled tubes in sulphuric acid and for unbundled in water, with water and hexane giving smaller up-shifts up to 30meV relative to air. For the bundled tubes especially, these shifts are remarkably large. Possibly they indicate that the liquids penetrate the bundles.

Also in Fig.2, we compare the shifts in bundled Hipco tubes under 2GPa pressure, using water as the pressure medium. A diamond-anvil cell was used, in the Zen configuration to give good pressure control at low pressures [18], with ruby luminescence as the pressure gauge. Three pressures were measured, up to 2GPa. The shift with pressure is almost orthogonal to the shift from air to water at ambient pressure. The shift in RBM frequency is to higher frequency, and greater for the larger tubes, as expected. The red-shift of the excitation resonance energy is substantial, and may be varying with chirality as well as diameter.

In Figure 3, some data is shown for the Carbolex tubes. Here, the identifiable points corresponded to the metallic families $2m + n = 33$ and 36. The shifts in the RBM frequency going from bundled in air to bundled in water are much larger, $\sim 5 – 7$ $cm^{-1}$, and the shifts in the excitation resonance energy are in the opposite direction to the semiconducting tubes, as observed by Maultzsch *et al.* [8] for different surfactants. For comparison we show also the calculated positions (in vacuum) given by Araujo *et al*. [16] and the experimental positions found by Strano *et al*. [17] for unbundled tubes in water. The behaviour here is very different from the semiconducting tubes in Fig.1, with large increases in the RBM frequencies relative to air, and with shifts in water and unbundled in water of the opposite sign to those in Fig.1.

**Conclusions**

A detailed discussion will be given when analysis of the data is complete. Meanwhile, the data of Figure 2 permit the following conclusions. The effect of the pressure medium is primarily on the excitation resonance energy and this has the confounding effect (described in Ref.1) on attempts to compare pressure experiments in different media at the same laser excitation energy. Different tubes are in resonance in the different media. It is clear that the effect of the medium is not to impose a "solvation pressure" as postulated previously [19]. Finally, while pressure imposes a closer contact and perhaps a stronger interaction between the medium and the nanotubes, this is not the principal effect of pressure. Most important, the data presented here show that the effects of pressure medium and of pressure may be clearly distinguished on these two-dimensional spectral plots. It will be possible to extract pressure coefficients for the RBM and G-bands for each $(m, n)$.

References

[1] D.J. Dunstan and A.J. Ghandour, *High-pressure studies of carbon nanotubes*, High Pres. Res. 29 (2009), pp 548-553.
[2] M. Hanfland, H. Beister and K. Syassen, *Graphite under pressure - equation of state and 1st-order Raman modes*, Phys. Rev. B39 (1989), pp 12598-12603.
[3] J.E. Proctor, E. Gregoryanz, K.S. Novoselov, M. Lotya, J.N. Coleman and M.P. Halsall, High-pressure Raman spectroscopy of grapheme, Phys. Rev. B80 (2009), 073408.
[4] A. Merlen, P. Toulemonde, N. Bendiab, A. Aouizerat, J.L. Sauvajol, G. Montagnac, H. Cardon, P. Petit and A. San Miguel, *Raman spectroscopy of open-ended single wall carbon nanotubes under pressure: effect of the pressure transmitting medium*, phys. stat. sol. (b)243 (2006), pp 690-699.
[5] A.L. Aguiar, E.B. Barros, R.B. Capaz, A.G. Souza Filho, P.T.C. Freire, J. Mendes Filho, D.Machon, C. Caillier, Y.A. Kim, H. Muramatsu, M. Endo and A. San Miguel, *Pressure-induced collapse in double-walled carbon nanotubes: chemical and mechanical screening effects*, J. Phys. Chem. C115 (2011), pp 5378-5384.
[6] S. You, M. Mases, I. Dobryden, A.A. Green, M.C. Hersam and A.V. Soldatov, *Probing structural stability of double-walled carbon nanotubes at high non-hydrostatic pressure by Raman spectroscopy*, High Pres. Res. 31 (2011), pp 186-190.
[7] H. Kataura, Y. Kumazawa, Y. Maniwa, I. Umeza, S. Suzuki, Y. Ohtsuka and Y. Achiba, *Optical properties of single-wall carbon nanotubes*, Synthetic Metals 103 (1999), pp 2555-2558.
[8] J. Maultzsch, H. Telg, S. Reich and C. Thomsen, *Radial breathing mode of single-walled carbon nanotubes: Optical transition energies and chiral-index assignment*, Phys. Rev. B72 (2005), 205438.
[9] H. Telg, J. Maultzsch, S. Reich and C. Thomsen, *First and second optical transitions in single-walled carbon nanotubes: a resonant Raman study*, phys. stat. sol. (b) 244 (2007), pp 4006-4010.
[10] S. Cambré, B. Schoeters, S. Luyckx, E. Goovaerts and W. Wenseleers, *Experimental observation of single-file water filling of thin single-wall carbon nanotubes down to chiral index (5,3)*, Phys. Rev. Lett 104 (2010), 207401.
[11] C. Fantini, A. Jorio, M. Souza, M.S. Strano, M.S. Dresselhaus and M.A. Pimenta, *Optical transition energies for carbon nanotubes from resonant Raman spectroscopy: Environment and temperature effects*, Phys. Rev. Lett. 93 (2004), 147406.
[12] H.D. Li, K.T. Yue, Z.L. Lian, Y. Zhan, L.X. Zhou, S.L. Zhang, Z.J. Shi, Z.N. Gu, B.B. Liu, R.S. Yang, H.B. Yang, G.T. Zou, Y. Zhang and S. Iijima, *Temperature dependence of the Raman spectra of single-wall carbon nanotubes*, Appl. Phys. Lett. 76 (2000), pp 2053-2055.
[13] M.J. O'Connell, S.M. Bachilo, C.B. Huffman, V.C. Moore, M.S. Strano, E.H. Haroz, K.L. Rialon, P.J. Bou, W.H. Noon, C. Kittrell, J. Ma, R.H. Hauge, R.B. Weisman and R.E. Smalley, *Band gap fluorescence from individual single-walled carbon nanotubes*, Science 297 (2002), pp 593-596.
[14] M. Paillet, P.Poncharal, A. Zahab, J.L. Sauvajol, J.C. Meyer and S. Roth, *Vanishing of the Breit-Wigner-Fano component in individual single-wall carbon nanotubes*, Phys. Rev. Lett. 94 (2005), 237401.
[15] P.T. Araujo, A. Jorio, M.S. Dresselhaus, K. Sato and R. Saito, *Diameter dependence of the dielectric constant for the excitonic transition energy of single-wall carbon nanotubes*, Phys. Rev. Lett. 103 (2009) 146802.
[16] P.T. Araujo, P.B.C. Presce, M.S. Dresselhaus, K. Sato, R. Saito and A. Jorio, *Resonance Raman spectroscopy of the radial breathing modes in carbon nanotubes*, Phys. E42 (2010), pp 1251-1261.
[17] M.S. Strano, S.K. Doom, E.H. Haroz, C. Kittrell, R.H. Hsuge and R.E. Smalley, *Assignment of (n, m) Raman and optical features of metallic single-walled carbon nanotubes*, Nano Letters 3 (2003), pp 1091-1096.

[18] N.W.A. van Uden and D.J. Dunstan, *Zen diamond-anvil low-pressure cell,* Rev. Sci. Instrum. 71 (2000), pp 4174-4176.

[19] J.R. Wood, M.D. Frogley, E.R. Meurs, A.D. Prins, T. Peijs, D.J. Dunstan and H.D. Wagner, *Mechanical response of carbon nanotubes under molecular and macroscopic pressures*, J. Phys. Chem. B103 (1999), pp 10388-10392.

**Figures and Captions**

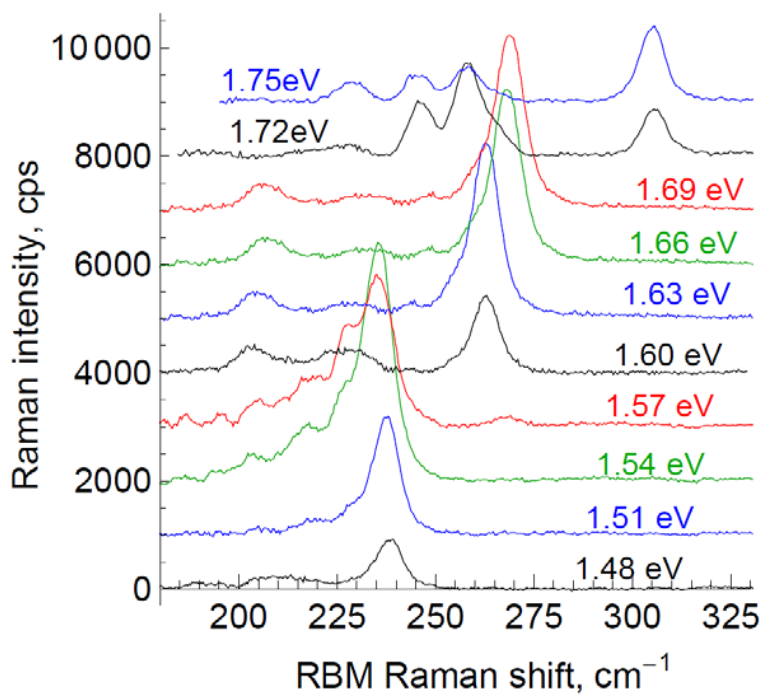

Figure 1 (Colour online): Ten spectra from the set obtained for Hipco semiconducting nanotubes, bundled, in water. The excitation energy for each spectrum is marked. The spectra are offset vertically for clarity.

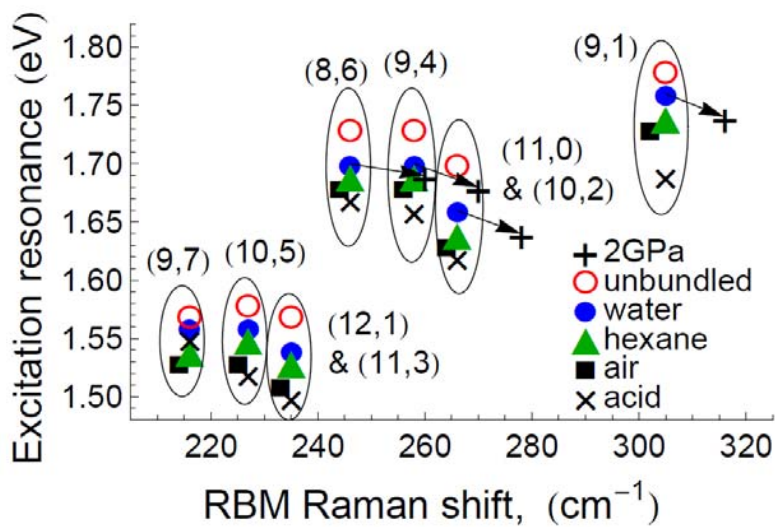

Figure 2 (Colour online): Each data point corresponds to the RBM Raman shift and the excitation resonance energy for Hipco semiconducting nanotubes with the chiral indices (*m*, *n*) marked. The squares are for dry nanotubes in air, the crosses (×) for sulphuric acid; the triangles for hexane. The solid circles are for bundled tubes in water and the open circles for unbundled in water with surfactant. Data taken under the pressure of 2GPa is shown by the crosses (+) for bundled nanotubes in water. The arrows show the shifts due to pressure.

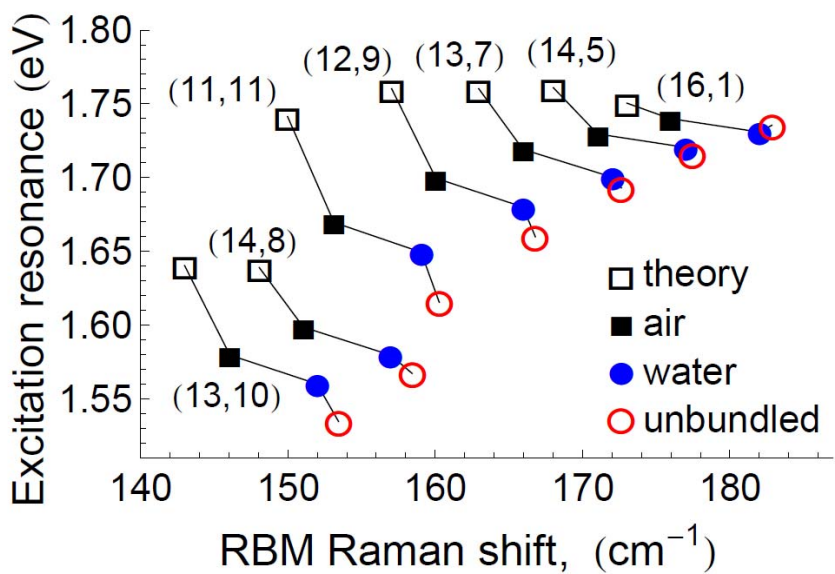

Figure 3 (Colour online): Each data point corresponds to the RBM Raman shift and the excitation resonance energy for Carbolex metallic nanotubes with the chiral indices (*m*, *n*) marked. The solid squares are for air and the solid circles for water. The open squares are the calculated positions from Araujo *et al.* [15] and the open circles are the experimental results of Strano *et al.* [17] for unbundled in water.